\documentclass[preprint,12pt]{elsarticle}

\usepackage{amsmath}
\usepackage{amssymb}
\usepackage{booktabs}
\usepackage{array}
\usepackage{url}
\usepackage{xcolor}
\usepackage{xspace}
\usepackage{graphicx}
\usepackage{placeins}
\graphicspath{{figures/}{./}}

\biboptions{numbers,sort&compress}

\newcounter{bla}

\journal{Computer Physics Communications}

\newcommand{\mlegs}{\textsc{MLegS}\xspace}

\begin{document}

\begin{frontmatter}

\title{\mlegs: A modern mapped Legendre spectral method solver for unbounded domains with parallelization}

\author[ntu]{Sangjoon Lee}%
\author[berkeley]{Jinge Wang}

\address[ntu]{School of Mechanical and Aerospace Engineering, Nanyang Technological University, \\Singapore 639798, Singapore}
\address[berkeley]{Department of Mechanical Engineering, University of California, Berkeley, \\Berkeley, California 94720, United States}

\begin{abstract}
\mlegs is an open-source Modern Fortran code package for solving linear and nonlinear time-dependent partial differential equations on radially unbounded domains. It builds on an earlier in-house prototype, recasting it as a modular solver with a unified interface for fields, transforms, and spectral operators. Within this framework, the numerical formulation maps $0\leq r<\infty$ onto a finite interval, where associated Legendre functions provide the radial basis, while Fourier series represent the periodic azimuthal and axial directions. The mapped radial basis represents radial infinity without imposing an artificial boundary condition at finite radius while incorporating the regularity required at $r=0$ according to azimuthal order. The implementation distributes the corresponding field layouts using the Message-Passing Interface (MPI), with strong- and weak-scaling measurements extending to 512 cores across four compute nodes. Beyond parallel performance, three examples assess the solver across diffusion, reaction--diffusion, and vortex dynamics: closed-form diffusion solutions establish spectral spatial convergence and the design orders of both semi-implicit integrators; a Fisher--Kolmogorov--Petrovsky--Piskunov front matches the linearized solution and propagates at the classical speed within its logarithmic correction; and vortex-pair simulations based on toroidal--poloidal decomposition yield a measured Crow growth rate $2.7\%$ above the theoretical prediction and demonstrate the reconnection of the vortex pair.
\end{abstract}


\end{frontmatter}

\section{Introduction}
\label{sec:introduction}

Localized dynamics in radially unbounded domains arise across several areas of physics, from vortices and reaction fronts to dispersive waves and quantum wave packets. Their governing equations are posed on infinite or semi-infinite intervals, but numerical discretizations conventionally replace those intervals with finite boxes. This replacement introduces an artificial outer boundary whose location, boundary condition, and possible motion become additional numerical choices that can affect the physical solution. Enlarging the box postpones rather than removes these choices and can devote most grid points to a region in which the solution magnitude is already negligible. By contrast, spectral representations on unbounded intervals offer a complementary strategy by incorporating infinity into the approximation space itself \cite{boyd1987rational,shen2009unbounded}. With suitable smoothness, decay, and endpoint compatibility, and with an appropriate map scale, such a representation can retain global high-order convergence without an explicitly imposed finite-radius boundary condition.

Several families realize this general strategy, each with its own advantages and disadvantages. Modified Legendre-rational functions provide half-line bases with useful approximation properties \cite{guo2001legendre}. Laguerre--Legendre compositions separate an exterior region from a bounded interior \cite{wang2009exterior}. Mapped Chebyshev or Hermite representations support fast or multidimensional formulations for particular operator classes \cite{sheng2020mappedchebyshev,guo2019hermite}. For these mapped approaches, a key practical consideration is the map scale, which dictates where collocation points concentrate; selecting or adapting this scale consequently remains an active topic \cite{xia2021scaling}. Ultimately, the diversity of these constructions highlights broader design criteria for a program library. Here, the central question is not whether an unbounded basis is universally superior to domain truncation, but whether its regularity, operator sparsity, transform cost, and coordinate geometry match the specific requirements of the target equations.

\begin{figure}[t]
\centering
\includegraphics[width=\textwidth]{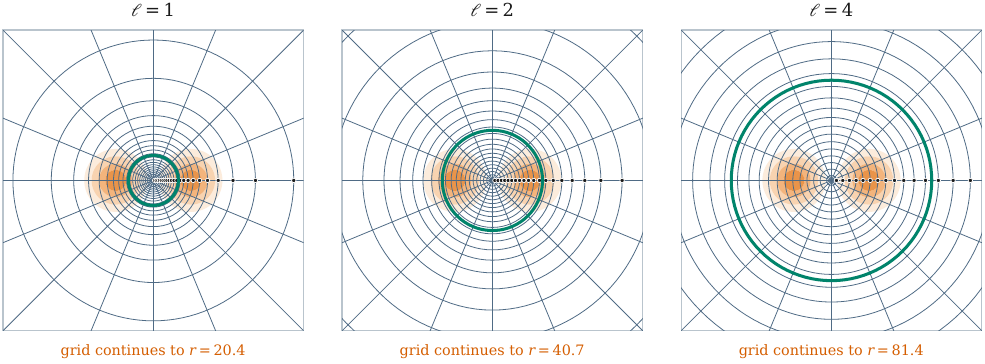}
\caption{The unbounded polar grid on which \mlegs\ discretizes a scalar, for $N_r=24$ radial and $N_\phi=16$ azimuthal nodes at three map scales $\ell$. The bold circle marks $r=\ell$; dots on the horizontal ray mark the radial nodes; shading marks a fixed pair of vortex cores. Each window spans $|x|,|y|\leq6$, while the grid itself continues to the radius quoted beneath each panel.}
\label{fig:grid}
\end{figure}

Guided by these criteria, \mlegs adopts the mapped associated Legendre construction introduced by Matsushima and Marcus \cite{matsushima1997unbounded}. This construction couples an algebraic map, which sends one endpoint to the polar or cylindrical origin and the other to radial infinity, with associated Legendre functions that supply the appropriate near-origin behavior for each azimuthal harmonic of order $m$. The coupling is particularly useful in polar and cylindrical geometry because regularity at $r=0$ depends on angular order, which can otherwise be obscured when the radial line is treated as an ordinary Cartesian interval. The map also provides direct control over the placement of radial resolution. Figure~\ref{fig:grid} shows the polar grid produced by \mlegs\ for $N_r=24$ radial and $N_\phi=16$ azimuthal nodes at three map scales $\ell$. As marked by the bold circle in each panel, exactly half the radial nodes lie inside $r=\ell$, and thus $\ell$ positions the high-resolution region to match the structure being computed \cite{lee2023mappedstability}. Beyond this scale, the nodes become more widely spaced with radius. In the cases shown, the outermost node lies between $r\approx20$ and $r\approx81$; increasing $N_r$ moves this outermost collocation point outward while improving radial resolution. The mapped representation thereby resolves a selected local scale while representing radial infinity without a finite-radius outer boundary condition.

Building on this numerical construction, the present work extends the earlier serial prototype into a reusable parallel implementation with a modular software structure and accompanying verification studies. 
This modernized software package has been primarily motivated by applications in vortex dynamics: aircraft wake vortices develop cooperative and elliptic instabilities before merger \cite{bristol2004cooperative}, while trailing-vortex models support related elliptical-instability mechanisms \cite{feys2016elliptical}. Within this application area, our framework has been used to address the linear stability of an isolated vortex with continuous eigenspectra \cite{lee2023mappedstability} and its nonmodal transient growth, including initiation by inertial particles \cite{lee2025transientgrowth}. Related weakly nonlinear theory, emphasizing selection rules and critical layers in triadic interactions, has also been investigated using the same numerical platform \cite{wang2026selectionrules}. However, the formulation has a broader scope than fluid mechanics, and the reaction--diffusion example is included to demonstrate this wider applicability.

From a software perspective, \mlegs belongs to an active ecosystem of open-source spectral tools that serve different computational settings. \textsc{Dedalus}\xspace features symbolic spectral PDE construction across multiple bases \cite{burns2020dedalus}; \textsc{Nektar++}\xspace supplies a high-order spectral/$hp$ element ecosystem \cite{moxey2020nektarpp}; \textsc{PyFR}\xspace offers cross-platform high-order incompressible-flow capability \cite{loppi2018pyfr}; and \textsc{spectralDNS}\xspace demonstrates concise high-performance Fourier algorithms for periodic turbulence \cite{mortensen2016spectraldns}. Other published programs address non-fluid time-dependent problems \cite{wang2010chebyshevtau,antoine2015gpelab}, while recent reports describe heterogeneous spectral solvers and spectral-element frameworks \cite{roccon2025flow36,massaro2024nekframework}. Collectively, these tools make different choices of bases and geometry, including bounded meshes and fully periodic configurations. \mlegs complements these programs by combining an origin-compatible mapped radial coordinate with natural support for unboundedness.

The remainder of the paper follows this progression from software design through numerical formulation and parallel performance to validation. Section~\ref{sec:structure} describes the package structure and modular programming, and Section~\ref{sec:formulation} presents the numerical formulation that drives the solver. Section~\ref{sec:parallel} then introduces the parallel decomposition and reports strong and weak scaling. Section~\ref{sec:examples} presents three practical examples together with their governing problems, computed outcomes, and validation results. Finally, Section~\ref{sec:conclusions} summarizes the findings.

\section{Code package structure}
\label{sec:structure}

\begin{figure}[t]
\centering
\includegraphics[width=\textwidth]{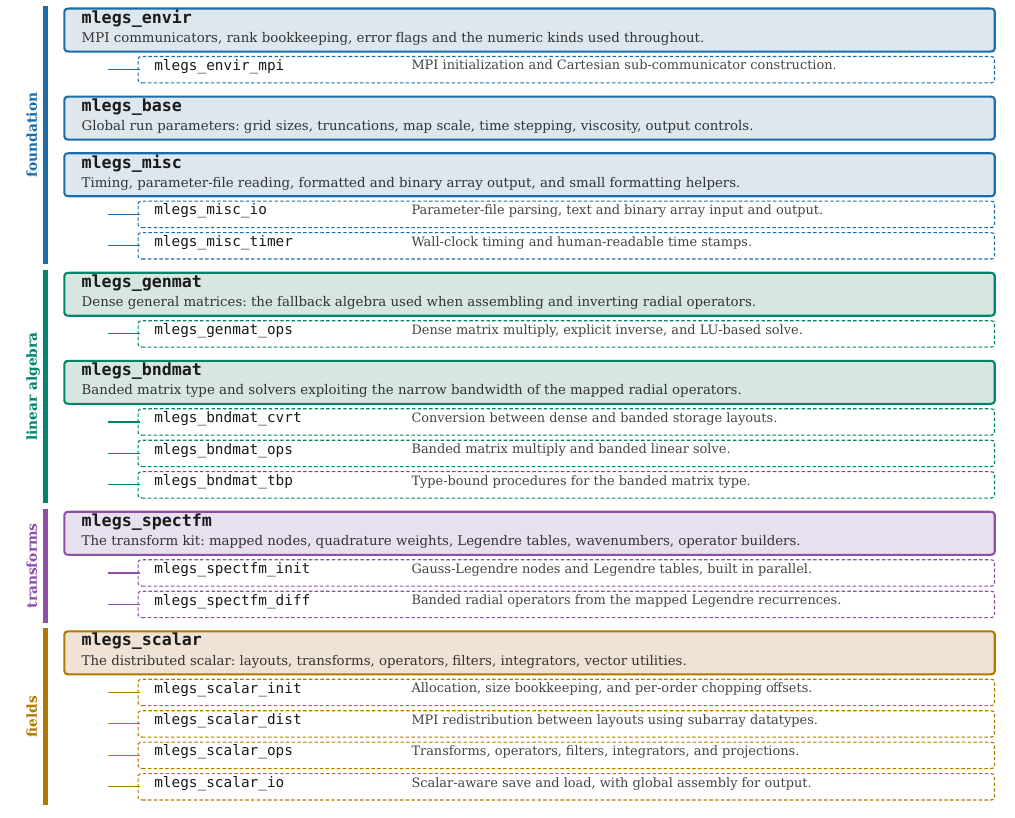}
\caption{Module and submodule hierarchy of \mlegs. Seven public modules (solid boxes) delegate their implementations to thirteen submodules (dashed boxes), organized in four layers with dependencies running upward.}
\label{fig:structure}
\end{figure}

Figure~\ref{fig:structure} summarizes the package: seven public modules (\texttt{src/modules}) delegate their implementations to thirteen submodules (\texttt{src/submodules}). Together they form a conceptual four-layer hierarchy with dependencies running upward. The foundation layer defines numeric kinds, the MPI environment and global run parameters. The linear-algebra layer provides dense-matrix routines and banded matrix types; the latter is central in practice because the mapped radial operators are narrow-banded. Next, the transform kit stores the mapped nodes, quadrature weights, associated Legendre tables and wavenumbers, and constructs radial operators from the mapped Legendre recurrences developed below. In the last layer, one distributed scalar type stores field data, and its module provides transforms, operators, filters, integrators and toroidal--poloidal utilities. Fourteen built-in application drivers (\texttt{src/apps}) complete the library, each illustrating a full call sequence for users and developers.

Two aspects of this organization distinguish the package from the original serial demonstration prototype:

\begin{itemize}
\item \textit{Modularization.} Each public interface is declared in a module, while each long numerical body resides in a separate submodule. This division keeps interfaces readable, lets the compiler check argument consistency throughout the package, and avoids recompiling dependent code unnecessarily. A user can therefore revise a transform, operator or integrator in its implementation submodule while leaving the surrounding interfaces intact.

\item \textit{Abstraction.} A field and a transform kit are derived types: each combines data with the procedures that act on it. The field type records its dimensions, distribution state, and physical or spectral representation, and the procedures that apply transforms and operators use that metadata.
Users apply operations through generic interfaces that take the field as an argument rather than by manipulating loop bounds or distributed array slices directly.
\end{itemize}

The library is built with an MPI-enabled Fortran 2008 compiler under Linux or another Unix-like system. The multiprecision package \textsc{FM}\xspace \cite{smith2011fm} and fast Fourier transform package \textsc{FFTE}\xspace \cite{takahashi2000ffte} are compiled from vendored sources, and \textsc{LAPACK}\xspace \cite{anderson1999lapack}, built from source in the same step, supplies the linear-algebra engine. A Python driver acts as the release gate: it checks the notebook and document inventory, builds and runs every documented application, verifies that an activated filter configuration changes the solution, and checks that malformed inputs are rejected. The next section develops the numerical foundation shared by these components.

\section{Numerical formulation}
\label{sec:formulation}

\subsection{Mapped Legendre representation}
\label{sec:rep}

We represent a three-dimensional scalar field $s(r,\phi,z,t)$ on $0\leq r<\infty$, $0\leq\phi<2\pi$, and $0\leq z<L_z$, with $\phi$ and $z$ periodic. Degenerate choices recover a one-dimensional radial problem or a two-dimensional polar problem. The algebraic map between $r$ and $x\in[-1,1)$, together with its inverse, is
\begin{equation}
x=\frac{r^2-\ell^2}{r^2+\ell^2},\qquad
r=\ell\sqrt{\frac{1+x}{1-x}},
\label{eq:map}
\end{equation}
where the user-selected map parameter $\ell>0$ places half of the Gauss--Legendre points within $r<\ell$. It is a resolution parameter rather than an outer radius, and is chosen to match the local structural scale of interest. For a real scalar, the truncated representation takes the form
\begin{equation}
s(r,\phi,z,t)=
\sum_{|k|<N_z^{c}}\;\sum_{|m|<N_\phi^{c}}\;\sum_{n=|m|}^{N_r^{c}-1}
\widehat{s}_{nmk}(t)\,
\bar{P}_n^m\!\left(x(r)\right)
\exp\!\left(\mathrm{i}m\phi+\mathrm{i}\kappa_k z\right),
\label{eq:expansion}
\end{equation}
where $\bar{P}_n^m$ denotes the normalized associated Legendre function of degree $n$ and order $m$, $\kappa_k=2\pi k/L_z$ is the axial wavenumber, and $m$ and $k$ are integers. The limits $N_r^{c}$, $N_\phi^{c}$ and $N_z^{c}$ truncate the radial, azimuthal and axial expansions, and the code and its input files call each limit a \emph{chop}. Each chop is bounded by the physical grid from which it is taken: $N_r^{c}\leq N_r$ for the $N_r$ radial collocation nodes, and $N_\phi^{c}\leq N_\phi/2+1$ and $N_z^{c}\leq N_z/2+1$ for the $N_\phi$ azimuthal and $N_z$ axial collocation nodes. Setting a chop below its bound discards the corresponding high modes and leaves headroom against quadratic aliasing (Section~\ref{sec:time}); unless noted otherwise, we keep the radial chop at its bound in this study. Because we consider $s$ to be real in the physical space, $\widehat{s}_{n,-m,-k}$ is the complex conjugate of $\widehat{s}_{nmk}$, and thus only coefficients with non-negative $m$ are stored. The constraint $n\geq|m|$ further reduces the retained radial count to $N_r^{c}-|m|$ at angular order $m$, making the active coefficient set triangular.

Under Eq.~\eqref{eq:map}, the associated Legendre equation becomes
\begin{equation}
\frac{\mathrm{d}}{\mathrm{d}r}\left(r\frac{\mathrm{d}P_{\ell,n}^m}{\mathrm{d}r}\right)
-\frac{m^2}{r}P_{\ell,n}^m
+\frac{4n(n+1)\ell^2r}{{(\ell^2+r^2)}^2}P_{\ell,n}^m=0,
\label{eq:mapped-legendre}
\end{equation}
writing $P_{\ell,n}^m(r):=\bar{P}_n^m(x(r))$. This is a Sturm--Liouville equation with weight $w(r)=4\ell^2r/{(\ell^2+r^2)}^2$, and thus $\{P_{\ell,n}^m\}_{n\geq|m|}$ forms a complete orthogonal basis set and supplies the sparse recurrences from which the radial operators are built \cite{matsushima1997unbounded,lee2023mappedstability}. A scalar analytic at the origin obeys the pole condition: its $m$th azimuthal harmonic behaves as $r^{|m|}$ times an even power series in $r$ as $r\rightarrow0$. Since $\bar{P}_n^m$ is $(1-x^2)^{|m|/2}$ times a polynomial in $x$, and both $x$ and $(1-x^2)^{1/2}/r$ are even functions of $r$, every basis function $P_{\ell,n}^m$ is $r^{|m|}$ times an even function of $r$, and thus each truncated harmonic in Eq.~\eqref{eq:expansion} meets the pole condition automatically. At the opposite endpoint, $x\rightarrow1$ represents $r\rightarrow\infty$, where the same factor $(1-x^2)^{|m|/2}$ makes every basis function of order $m\neq0$ decay as $O(r^{-|m|})$, a decay that any truncation at $N_r^c$ preserves. The axisymmetric functions are the exception: $\bar{P}_n^0$ tends to a nonzero constant at $x=1$, and thus an $m=0$ component has a free far-field level. The package supplies an operation that removes it, adjusting the regular coefficients so that the limit vanishes wherever the physics requires decay.

Differential operations are formed in coefficient space. For instance, writing $\widehat{s}_{mk}$ for the coefficient vector of the $(m,k)$ harmonic of $s$, the Laplacian takes the form
\begin{equation}
\widehat{\nabla^2s}_{mk}=\left(\mathsf{D}_m-\kappa_k^2\mathsf{I}\right)\widehat{s}_{mk},
\label{eq:laplacian}
\end{equation}
where $\mathsf{I}$ denotes the identity and $\mathsf{D}_m$ the matrix form of the horizontal Laplacian $\nabla_h^2\equiv r^{-1}\partial_r(r\partial_r)+r^{-2}\partial_\phi^2$, which on the $(m,k)$ harmonic reduces to $r^{-1}\partial_r(r\partial_r)-m^2r^{-2}$, just as the second axial derivative reduces to the multiplier $-\kappa_k^2$. The recurrences of Eq.~\eqref{eq:mapped-legendre} couple a degree only to its immediate neighbors, and thus $r\,\partial_r$ is tridiagonal; the second-order operator applies that coupling twice, and hence $\mathsf{D}_m$ is pentadiagonal: for every $n\geq|m|$,
\begin{equation}
\begin{aligned}
\ell^2\,\widehat{\nabla_h^2s}_{nmk}
={}&-\frac{(n-|m|-1)(n-|m|)(n-2)(n-1)}{(2n-3)(2n-1)}\,\widehat{s}_{(n-2)mk}\\
&+\frac{2n(n-1)(n-|m|)}{2n-1}\,\widehat{s}_{(n-1)mk}\\
&-\frac{2n(n+1)(3n^2+3n-m^2-2)}{(2n-1)(2n+3)}\,\widehat{s}_{nmk}\\
&+\frac{2(n+1)(n+2)(n+|m|+1)}{2n+3}\,\widehat{s}_{(n+1)mk}\\
&-\frac{(n+2)(n+3)(n+|m|+1)(n+|m|+2)}{(2n+3)(2n+5)}\,\widehat{s}_{(n+2)mk},
\end{aligned}
\label{eq:del2h-entries}
\end{equation}
under the convention $\widehat{s}_{nmk}\equiv0$ for $n<|m|$ \cite{matsushima1997unbounded,lee2023mappedstability}. The entries are rational in $n$ and $m$ and scale as $\ell^{-2}$; those written here act on the unnormalized $P_n^m$, and the code rescales each by a ratio of the normalization factors defined below.

Bandedness makes these operators affordable: the banded product with Eq.~\eqref{eq:laplacian} costs $O(N_r^{c})$ rather than the $O({(N_r^{c})}^2)$ of a full matrix product, and the banded solve $O(N_r^{c})$ rather than $O({(N_r^{c})}^3)$ through the banded LU factorization and back substitution of the \textsc{LAPACK}\xspace linear-algebra engine. In addition, storage falls from ${(N_r^{c})}^2$ to five diagonals. The same construction renders $r\,\partial_r$ tridiagonal, while Helmholtz, powered Helmholtz for $p\in\{4,6,8\}$ and their inverses follow by shifting and composing $\mathsf{D}_m$.

Inversion of the Laplacian requires one additional consideration. As all basis functions remain bounded as $r\rightarrow\infty$, no combination of them can represent the logarithmic far field that a two-dimensional Poisson solution acquires when its source has a nonzero integral \cite{matsushima1997unbounded}. A scalar therefore carries one extra real coefficient $c$ beyond Eq.~\eqref{eq:expansion}, multiplying the analytic proxy $\Lambda(r)=\ln(r^2+\ell^2)$. Although $\Lambda$ itself lies outside the basis, its horizontal Laplacian falls inside it, in three axisymmetric terms,
\begin{equation}
\nabla_h^2\Lambda=\frac{1}{\ell^2}\left(\frac{4}{3}P_0^0-2P_1^0+\frac{2}{3}P_2^0\right),
\label{eq:logproxy}
\end{equation}
where the unnormalized $P_n^0$ are evaluated at $x(r)$. A forward Laplacian adds these three coefficients to the $m=k=0$ harmonic block and clears the logarithmic part of the result. Inversion offers two variants differing only within that block: one prescribes the amplitude $c$, replacing the first row of $\mathsf{D}_0$ by that constraint, and the other solves for it, replacing the first column by the image in Eq.~\eqref{eq:logproxy} so that the unknown multiplying $P_{\ell,0}^0$ becomes $c$. All other harmonics remain ordinary banded solves.

The normalization in Eq.~\eqref{eq:expansion} is $\bar{P}_n^m=N_n^mP_n^m$, with
\begin{equation}
N_n^m=\sqrt{\frac{(2n+1)(n-|m|)!}{2\,(n+|m|)!}},
\label{eq:norm}
\end{equation}
which absorbs the factorials left by the Gauss--Legendre quadrature. Evaluated as written, Eq.~\eqref{eq:norm} is unusable in double precision: at $n=200$ and $|m|=100$ the factorial $(n+|m|)!$ reaches $10^{614}$ against an overflow threshold near $10^{308}$, and the recurrence generating $P_n^m$ carries a leading double factorial $(2|m|-1)!!$ that overflows on its own past order $|m|\approx150$. The code therefore stores $\ln N_n^m$, accumulated from the ratio $2\ln(N_n^m/N_{n-1}^m)=\ln[(2n+1)(n-|m|)]-\ln[(2n-1)(n+|m|)]$ in which only quantities of order unity appear, and builds the tables in \textsc{FM}\xspace multiprecision before rounding to double. Operator entries are scaled the same way, by $\exp(\ln N_j^m-\ln N_i^m)$ rather than by a ratio of factorials. The multiprecision arithmetic cost is paid only once at start-up and distributed over the process grid; every subsequent operation runs in double precision.

\subsection{Vector fields}
\label{sec:vector}

The operators above act on a single scalar, and the straightforward extension to a three-dimensional vector field is to carry its three primitive components as three scalars. For a solenoidal field, however, one of the three is redundant, and \mlegs\ instead offers a two-component representation through the toroidal--poloidal projection, which also removes any gradient contribution the input field may carry.

With $\hat{\boldsymbol{e}}_z$ as the reference direction, a sufficiently rapidly decaying solenoidal $\boldsymbol{V}$ is expressed as
\begin{equation}
\boldsymbol{V}=\nabla\times(\psi\hat{\boldsymbol{e}}_z)
+\nabla\times\nabla\times(\chi\hat{\boldsymbol{e}}_z),
\label{eq:tp}
\end{equation}
where $\psi$ and $\chi$ denote the toroidal and poloidal potentials, respectively \cite{matsushima1997unbounded,lee2023mappedstability}. We write the map from a field to its two potentials as $\mathcal{P}(\boldsymbol{V})\equiv(\psi,\chi)$. Three components thus become two scalars, each of which the mapped Legendre machinery already handles, and no separate divergence constraint remains.

The potentials are recovered from the axial component of Eq.~\eqref{eq:tp} and of its curl, each a two-dimensional Poisson problem in the plane, as follows:
\begin{equation}
\nabla_h^2\chi=-V_z,\qquad
\nabla_h^2\psi=-(\nabla\times\boldsymbol{V})_z,
\label{eq:tp-poisson}
\end{equation}
where $\nabla_h^2$ denotes the Laplacian in $(r,\phi)$, inverted with the banded operators of Section~\ref{sec:rep}. In the code, $\psi$ follows directly from tabulated projections and $\chi$ from one banded inversion, and thus evaluating $\mathcal{P}$ costs one elliptic solve per harmonic. Its inverse is the differentiation in Eq.~\eqref{eq:tp}. The construction is not unconditional: solving Eq.~\eqref{eq:tp-poisson} by convolution against the plane Green's function $(2\pi)^{-1}\ln r$ converges only if $V_r$ and $V_\phi$ decay faster than $r^{-1}$ and $V_z$ faster than $r^{-2}$, which for a velocity field also guarantees finite kinetic energy over the plane \cite{lee2023mappedstability}. A vortex with net circulation violates this and is admitted through the logarithmic proxy of Eq.~\eqref{eq:logproxy}. The residual gauge freedom, namely the addition of an arbitrary function of $z$ to either potential, is removed by requiring $\psi,\chi\rightarrow0$ as $r\rightarrow\infty$, making $\mathcal{P}$ linear and invertible on that class.

Two properties of this projection are central. The evolution equations for $\psi$ and $\chi$ follow from the axial curl and curl--curl of the momentum equation, both of which annihilate a gradient, and thus the pressure term drops out and no pressure Poisson solve, often the most computationally intensive step of a Navier--Stokes solver, is needed. The projection also commutes with the Laplacian, and thus a stiff dissipative operator acts on $\psi$ and $\chi$ independently, making a vector problem no more expensive to treat semi-implicitly than two scalars.

Origin regularity transfers as well, since potentials of azimuthal order $m$ expanded in the mapped basis behave as $O(r^{|m|+2j})$ for non-negative integer $j$, which makes the reconstructed velocity satisfy the centerline conditions automatically, including the derivative constraints that a pole condition imposed on the primitive components does not supply \cite{lee2023mappedstability}.

\subsection{Time integration}
\label{sec:time}

Every integrator in the package advances a scalar written in the archetypal split form
\begin{equation}
\frac{\partial s}{\partial t}=\mathcal{A}(s,t)+\mathcal{L}s,
\label{eq:split}
\end{equation}
in which $\mathcal{A}$ collects the problem-specific nonstiff contributions that a driver supplies (advection, reaction or forcing, evaluated pseudospectrally in physical space), and $\mathcal{L}$ is the stiff linear operator the library provides,
\begin{equation}
\mathcal{L}s=\nu\nabla^2s-\nu_h(-\nabla^2)^{p/2}s,
\qquad p\in\{4,6,8\},
\label{eq:dissipation}
\end{equation}
with $\nu_h\geq0$. The two coefficients play different roles: the physical viscosity $\nu$ is a property of the problem and is set to match its physical dissipation; the hyperviscosity $\nu_h$ is a numerical apparatus, present only to drain the energy that a truncated spectral representation accumulates at its smallest retained scales, and its steep $(-\nabla^2)^{p/2}$ weighting is chosen so that it acts there and as little as possible on the resolved dynamics. The split makes the semi-implicit schemes affordable: $\mathcal{L}$ is a Fourier multiplier in the two periodic directions and a banded matrix in the radial one, and thus treating it implicitly costs one inverse Helmholtz or powered Helmholtz application per step rather than a global inversion, while $\mathcal{A}$ is never linearized.

Four time integration schemes are provided: forward Euler applied to both terms; forward Euler for $\mathcal{A}$ with backward Euler for $\mathcal{L}$ (FE--BE); the second-order explicit Adams--Bashforth pair; and Adams--Bashforth with Crank--Nicolson (AB--CN),
\begin{equation}
\left(\mathsf{I}-\frac{\Delta t}{2}\mathcal{L}\right)s^{j+1}
=\left(\mathsf{I}+\frac{\Delta t}{2}\mathcal{L}\right)s^j
+\frac{\Delta t}{2}\left(3\mathcal{A}^j-\mathcal{A}^{j-1}\right).
\label{eq:abcn}
\end{equation}
The implicit step calls the inverse Helmholtz or powered Helmholtz operator in spectral space, removing the diffusive restriction but not the advective one. Second-order runs need a consistent previous state, for which the examples use Richardson extrapolation from FE--BE substeps.

Quadratic products are formed pseudospectrally, and aliasing in each direction is controlled by the margin between its chop and its collocation count: the package applies the conventional two-thirds mask in the periodic azimuthal and axial directions, and a radial margin is available by setting $N_r^{c}$ below $N_r$. The runs here leave no radial margin, and thus a radial product can fold energy from unresolved degrees into retained ones. Controlling the resulting high-mode tail is what the dissipation is for. A fixed $\nu_h$ is simple and enlarges the useful integration interval, but excessive values alter the resolved dynamics, and a steep hyperviscous operator can create a bottleneck near the truncation scale \cite{frisch2008hyperviscosity}. The package therefore also provides an optional filter, which multiplies every coefficient above a normalized cutoff $\zeta_c$ by $\exp[-\epsilon(\zeta-\zeta_c)^8/(1-\zeta_c)^8]$, where $\zeta$ is the largest of the radial, azimuthal and axial mode indices, each normalized to its own truncation. Its amplitude $\epsilon$ is not fixed but follows the fraction of coefficient energy lying above the cutoff: the filter stays inactive until that fraction first exceeds a target $\tau$ (by default $2\times10^{-2}$, i.e., $2\%$ of the total spectral energy), strengthens as it passes $\tau$, and saturates once it reaches $2\tau$. The response is bounded and relaxed across steps, and each field has its own state. Selective damping of the highest modes is the rationale behind spectral vanishing viscosity \cite{karamanos2000svv}, but the present implementation is a modal filter applied between steps rather than an added viscous term.

\section{Parallelization and performance}
\label{sec:parallel}

\subsection{Decomposition and transposes}
\label{sec:decomposition}

The decomposition follows from a single constraint: a spectral transform along a coordinate is a global operation along that coordinate, and thus every rank performing it must hold the whole line of data. With three transformable directions and no direction privileged, \mlegs\ uses a two-dimensional pencil decomposition: the ranks are arranged on a Cartesian process grid of size $p_1\times p_2$, and at any moment exactly one array axis is undistributed while the other two are spread over the two grid dimensions. Each scalar records, for every axis, which of the two sub-communicators owns it, with a null entry marking the local axis. The fully physical state \texttt{PPP} (in the order of radial, azimuthal and axial directions) keeps the azimuthal axis whole and distributes the radial and axial axes; the fully spectral state \texttt{FFF} keeps the axial axis whole instead.

\begin{figure}[t]
\centering
\includegraphics[width=\textwidth]{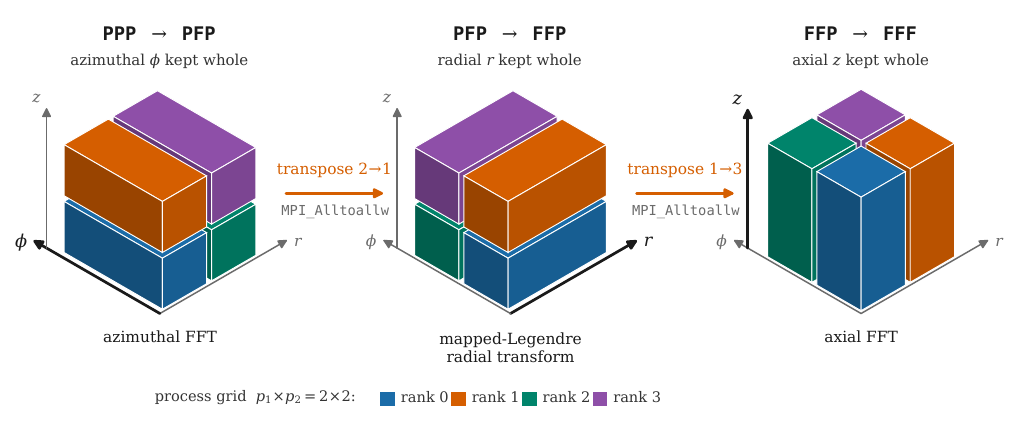}
\caption{Pencil decomposition and the transpose sequence of a forward transform, for a $2\times2$ process grid. Labels give the state string $(r,\phi,z)$ before and after the transform performed in that layout, with \texttt{P} and \texttt{F} for physical and spectral representation. The bold axis is the one kept whole on every rank; colors denote ranks.}
\label{fig:decomposition}
\end{figure}

Figure~\ref{fig:decomposition} follows a scalar array through a complete physical-to-spectral transform. The azimuthal Fourier transform is taken first, in the layout that already keeps $\phi$ whole. A transpose then exchanges the azimuthal and radial axes so that each rank owns complete radial lines, and the mapped Legendre transform of Section~\ref{sec:rep} (a dense matrix product along $r$) is applied locally. A second transpose exchanges the radial and axial axes, and the axial Fourier transform completes the sequence. Rank colors are held fixed across the three panels to make the invariant visible: a rank keeps its position on the process grid throughout, and what changes at each transpose is which axes that position indexes. The reverse order returns a field to physical space, and because the intermediate states \texttt{PFP} and \texttt{FFP} are addressable, a driver may hold a field in a mixed representation when an operator calls for one. Layout is likewise not frozen between transforms: the radial differentiation routines transpose to whole radial lines when they need them and transpose back on exit, and thus a caller sees only the state it asked for.

Apart from a few small global reductions, transposing is the only communication in the time loop. The redistribution routine builds MPI subarray datatypes describing the blocks each rank must send and receive, and performs the exchange with a single \texttt{MPI\_Alltoallw} within the relevant sub-communicator. Two properties matter. First, the datatypes let MPI gather strided blocks directly out of the array, and thus no packing buffer is needed and no globally replicated field is formed. Second, because only one grid dimension participates in each transpose, the all-to-all runs over $p_1$ or $p_2$ ranks rather than all $p_1p_2$ of them. Two-dimensional problems degenerate to a slab decomposition, $p_2=1$.

\subsection{Scaling}
\label{sec:scaling}

Scaling was measured on the Anvil cluster, whose compute nodes carry two 64-core AMD EPYC 7763 processors and 256\,GB of memory \cite{rcac2026anvil}, using GNU Fortran 11.2.0 with Open MPI 4.1.6 and the vendored \textsc{FM}\xspace (v1.4), \textsc{FFTE}\xspace (v7.0) and \textsc{LAPACK}\xspace (v3.12.0) builds.

The benchmark is a production calculation rather than a synthetic kernel. We simulate the co-rotating $q$-vortex pair of Section~\ref{sec:ex3} (two vortices at $(x,y)=(\pm2,0)$ with $q=1$ in a uniform axial background $U_z=-0.5$, at $\ell=4$, $L_z=2\pi$ and the Reynolds number $\mathrm{Re}\equiv\Gamma/\nu=6.3\times10^4$, where $\Gamma$ is the circulation of one vortex, seeded with a deterministic perturbation of amplitude $10^{-10}$), advanced by the same Richardson extrapolation start-up and AB--CN loop that the shipped three-dimensional application uses. The evolved state is the toroidal--poloidal pair $(\psi,\chi)$ on a $256\times256\times256$ grid with $N_r^{c}=256$, $|m|\leq128$ and $|k|\leq128$, which the triangular truncation of Eq.~\eqref{eq:expansion} reduces to about $6.3$ million retained complex coefficients per potential. Dissipation is the full operator of Eq.~\eqref{eq:dissipation}, with hyperviscosity of order $p=8$ at $\nu_h=5\times10^{-7}$ and the bounded filter enabled. Each step costs two implicit solves, one per potential, and one nonlinear evaluation; the latter includes the full transform round trip.

Strong scaling advances this fixed problem 30 steps at $\Delta t=10^{-2}$; weak scaling grows the axial size in proportion to the core count so that the work per rank is fixed. Field output is disabled in both, and thus the timings measure computation and communication rather than the parallel file system. Table~\ref{tab:scaling} reports both, spanning a single node to four nodes.

\begin{table}[t]
\centering
\caption{Strong and weak scaling, in seconds over 30 steps with field output disabled. Strong scaling advances one fixed $256\times256\times256$ problem, with efficiency relative to the 16-rank run; weak scaling grows the axial size with the core count at fixed work per rank.}
\label{tab:scaling}
\small
\begin{tabular*}{\textwidth}{@{\extracolsep{\fill}}ccccccccc@{}}
\toprule
& & \multicolumn{5}{c}{strong scaling, fixed $256^3$} & \multicolumn{2}{c}{weak scaling}\\
\cmidrule(lr){3-7}\cmidrule(lr){8-9}
nodes & ranks & total & speedup & efficiency & implicit & nonlinear & $N_z$ & total\\
\midrule
1 & 16  & 410.1 & 1.00  & ---     & 253.5 & 135.0 & ---  & ---\\
1 & 32  & 221.2 & 1.85  & $93\%$  & 135.1 &  73.8 & ---  & ---\\
1 & 64  & 115.3 & 3.56  & $89\%$  &  68.2 &  39.7 & ---  & ---\\
1 & 128 &  78.9 & 5.20  & $65\%$  &  44.8 &  27.7 & 64   & 21.3\\
2 & 256 &  42.6 & 9.64  & $60\%$  &  23.1 &  14.3 & 128  & 23.0\\
4 & 512 &  30.4 & 13.51 & $42\%$  &  13.2 &  10.8 & 256  & 30.4\\
\bottomrule
\end{tabular*}
\end{table}

Several features are worth drawing out. Within a single socket pair the code scales well, retaining $89\%$ efficiency at 64 ranks, but the step to 128 ranks, which fills the node, yields only $1.46\times$, a signature of memory-bandwidth saturation rather than of communication. Crossing from one node to two then recovers a speedup of $1.85\times$, because the additional node brings its own bandwidth with it; the subsequent step to four nodes yields $1.40\times$, by which point the per-rank problem has become small enough that the transposes likely dominate.

Weak scaling efficiency is $93\%$ at two nodes and $70\%$ at four. Transform-kit initialization, which is the part of the calculation most obviously amenable to embarrassing parallelism, falls from $2.20$ to $0.71$\,s across the strong-scaling range and is negligible against the time-stepping cost in any production run. Throughout, the implicit solves remain the single largest phase.

\section{Examples}
\label{sec:examples}

\subsection{Example 1: diffusion problems with exact solutions}
\label{sec:ex1}

\paragraph{Problem}
The first example collects three linear or manufactured problems whose solutions are known in closed form. Together they isolate the accuracy of the mapped representation and its operators, the order of the two semi-implicit integrators, and the treatment of the origin for a field that is not axisymmetric. Because the exact answers are available at every instant, these are the problems that fix what the discretization can and cannot do before any nonlinear calculation is attempted.

\paragraph{Governing equations and initial conditions}
The spatial accuracy of the representation is measured on two static fields chosen to separate the two distinct error mechanisms. A finite combination of mapped Legendre modes is represented exactly and thus cannot expose a convergence rate; we therefore use a rational field and a Gaussian field,
\begin{equation}
s_A=\frac{r\cos\phi}{r^2+a^2}+\frac{r^2\cos 2\phi}{{(r^2+a^2)}^2},
\qquad
s_B={\left(\frac{r}{\sigma}\right)}^{2}\exp\left(-\frac{r^2}{2\sigma^2}\right)\cos 2\phi,
\label{eq:cases}
\end{equation}
with $a=1.7$ and $\sigma=1$, both possessing closed-form Laplacians; $s_A$ has $m=1$ and $m=2$ content, while $s_B$ has only $m=2$. The essential difference is analyticity in the mapped coordinate: each harmonic of $s_A$ is $(1-x^2)^{|m|/2}$ times a rational function of $x$ with poles off $[-1,1]$, whereas the Gaussian tail of $s_B$ maps to a nonanalytic point at $x=1$.

Temporal accuracy is measured on the axisymmetric vorticity equation $\partial_t\omega=\nu(\partial_{rr}+r^{-1}\partial_r)\omega$ with unit circulation. Its exact solution is the classical Oseen family, the long-time attractor of two-dimensional vorticity at fixed circulation \cite{gallay2005oseen}
\begin{equation}
\omega(r,t)=\frac{1}{2\pi\sigma_0^2 B(t)}\exp\left[-\frac{r^2}{2\sigma_0^2B(t)}\right],
\qquad B(t)=1+\frac{2\nu t}{\sigma_0^2},
\label{eq:oseen}
\end{equation}
started from $B=1$ with $\sigma_0=1$, $\nu=0.1$ and $\ell=1$.

Origin regularity is tested with the manufactured heat mode
\begin{equation}
s_m(r,\phi,t)=A\,B(t)^{-(m+1)}\left(\frac{r}{\sigma_0}\right)^m
\exp\left[-\frac{r^2}{2\sigma_0^2B(t)}\right]\cos(m\phi),
\label{eq:gaussian-m}
\end{equation}
with the same $\sigma_0$, $\nu$ and $B(t)$ as Eq.~\eqref{eq:oseen}, which makes it satisfy $\partial_t s_m=\nu\nabla^2 s_m$ exactly; it is initialized at $t=0$, and the choice $m=2$ makes $s_m=O(r^2)$ at the origin with zero azimuthal mean. In every case the only condition imposed at large radius is decay, which the representation supplies for $m\neq0$; no outer boundary condition exists to be specified. Localized rotating two-dimensional structures \cite{grahameagle2002rotating} motivate non-axisymmetric tests of this kind, although Eq.~\eqref{eq:gaussian-m} is a manufactured mode rather than a reproduced solution.

\begin{figure}[t]
\centering
\includegraphics[width=\textwidth]{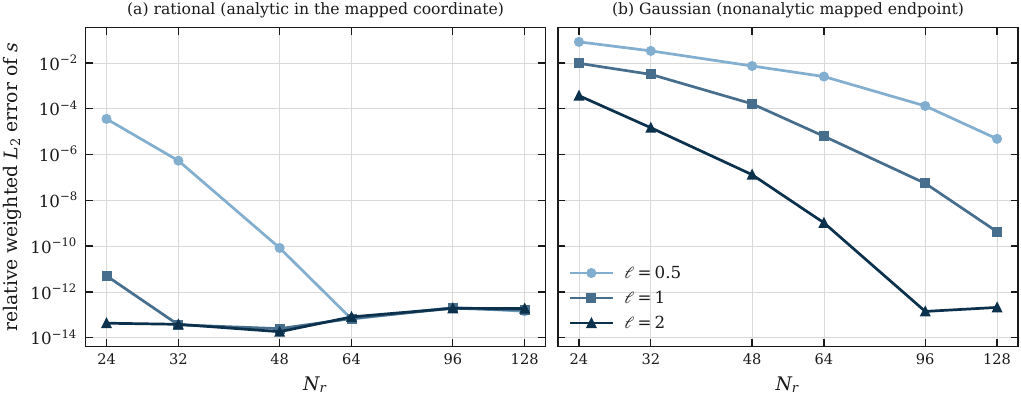}
\caption{Relative weighted $L_2$ representation error of $s$ against radial truncation, for three map scales. (a) The rational field of Eq.~\eqref{eq:cases}, analytic in the mapped coordinate. (b) The Gaussian field, whose tail maps to a nonanalytic endpoint.}
\label{fig:spatial}
\end{figure}

\paragraph{Results and validation}
Figure~\ref{fig:spatial} shows the two spatial-convergence families, and the behavior is the signature of a spectral representation rather than a fixed-order one: the error falls rapidly until it meets the double-precision round-off floor of $O(10^{-13})$. For the rational field, analytic in the mapped coordinate, that happens almost immediately; at $\ell=1$ the relative representation error is already $3.7\times10^{-14}$ by $N_r=32$. The Gaussian field, whose tail maps to a nonanalytic endpoint, converges more slowly but still superalgebraically: at $\ell=2$ it passes through $3.7\times10^{-4}$, $1.3\times10^{-7}$ and $1.0\times10^{-9}$ at $N_r=24$, $48$ and $64$ before reaching $1.4\times10^{-13}$ at $N_r=96$, its effective algebraic order rising with each refinement. In addition to the resolution check in terms of $N_r$, the map scale needs attention: at $N_r=64$ the Gaussian error improves from $2.5\times10^{-3}$ at $\ell=0.5$ to $1.0\times10^{-9}$ at $\ell=2$, which shows that $\ell$ must be chosen to match the problem's local dynamical scale.

\begin{figure}[t]
\centering
\includegraphics[width=\textwidth]{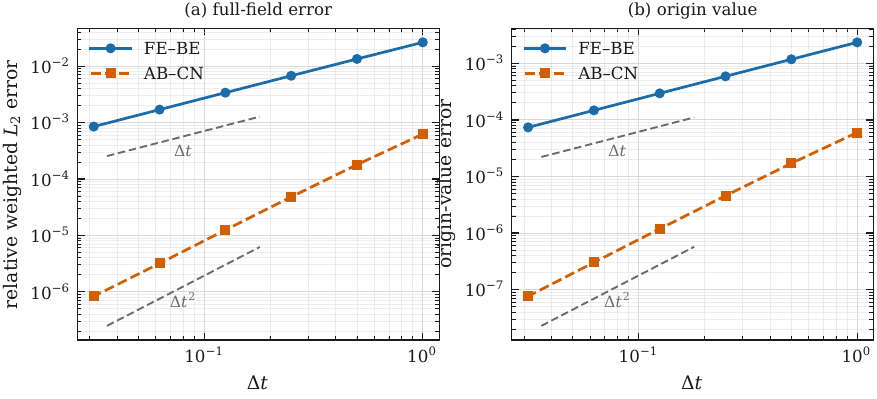}
\caption{Temporal convergence of the two semi-implicit integrators on the exact Oseen solution, with the spatial error held below the temporal error. (a) Relative weighted $L_2$ error of the full field. (b) Error in the origin value. Dashed guides indicate first- and second-order reference slopes.}
\label{fig:temporal}
\end{figure}

The sweep of Table~\ref{tab:oseen} and Fig.~\ref{fig:temporal} demonstrates accuracy in time while holding the spatial error far below the temporal one. Observed orders $p_t$ approach the design values, reaching $0.999$ for FE--BE and $1.938$ for AB--CN at the smallest step, the latter after peaking at $1.970$; because the nonlinear term vanishes identically here, FE--BE degenerates to backward Euler and AB--CN to Crank--Nicolson, and thus first and second order are what should be recovered. Before use, the driver was required to reproduce the origin values of the repository's first-order time-integration tutorial at its coarser resolution, which it does to relative differences of at most $2\times10^{-13}$. The integral invariants are more instructive than the norms. We track the circulation and the second radial moment,
\begin{equation}
\Gamma(t)=\int_0^{2\pi}\!\!\int_0^{\infty}\omega\,r\,\mathrm{d}r\,\mathrm{d}\phi,
\qquad
M_2(t)=\int_0^{2\pi}\!\!\int_0^{\infty}r^2\,\omega\,r\,\mathrm{d}r\,\mathrm{d}\phi,
\label{eq:invariants}
\end{equation}
both evaluated with the mapped quadrature; for Eq.~\eqref{eq:oseen} their exact values are $\Gamma=1$ and $M_2=2\sigma_0^2B(t)$, the latter equal to $6$ at $t_f=10$. At the smallest step AB--CN returns $\Gamma=0.9999936$ and $M_2=5.9551$, and neither moves in the digits shown once $\Delta t\leq0.125$, even though the field error is still falling by a factor of four per halving. What limits them is spatial rather than temporal. The $r^2$ weighting places the mass of $M_2$ in the far field, where the computed solution is least accurate, and thus the residual $0.75\%$ is spatial error that no reduction of $\Delta t$ can remove. Circulation, weighted only by $r$, samples the well-resolved core instead and lands within $6.4\times10^{-6}$ of unity. A purely temporal study would not reveal that these diagnostics are already spatially limited at this resolution.

\begin{table}[t]
\centering
\caption{Temporal convergence for the Oseen problem at $t_f=10$ with $N_r=96$. Errors are relative to the exact solution; $p_t$ is the observed order across adjacent halvings of the full six-step sweep, of which three steps are listed. The circulation $\Gamma$ and second radial moment $M_2$ of Eq.~\eqref{eq:invariants} are absolute values, against exact $\Gamma=1$ and $M_2=6$.}
\label{tab:oseen}
\small
\begin{tabular*}{\textwidth}{@{\extracolsep{\fill}}lccccc@{}}
\toprule
scheme & $\Delta t$ & rel.\ $L_{2,w}$ & $p_t$ & $\Gamma$ & $M_2$\\
\midrule
FE--BE & 1.0     & $2.646\times10^{-2}$ & ---   & 1.0000031 & 6.0215\\
       & 0.25    & $6.757\times10^{-3}$ & 0.990 & 0.9999995 & 5.9964\\
       & 0.03125 & $8.498\times10^{-4}$ & 0.999 & 0.9999945 & 5.9618\\
\midrule
AB--CN & 1.0     & $6.271\times10^{-4}$ & ---   & 0.9999945 & 5.9615\\
       & 0.25    & $4.810\times10^{-5}$ & 1.898 & 0.9999936 & 5.9552\\
       & 0.03125 & $8.309\times10^{-7}$ & 1.938 & 0.9999936 & 5.9551\\
\bottomrule
\end{tabular*}
\end{table}

\begin{figure}[t]
\centering
\includegraphics[width=\textwidth]{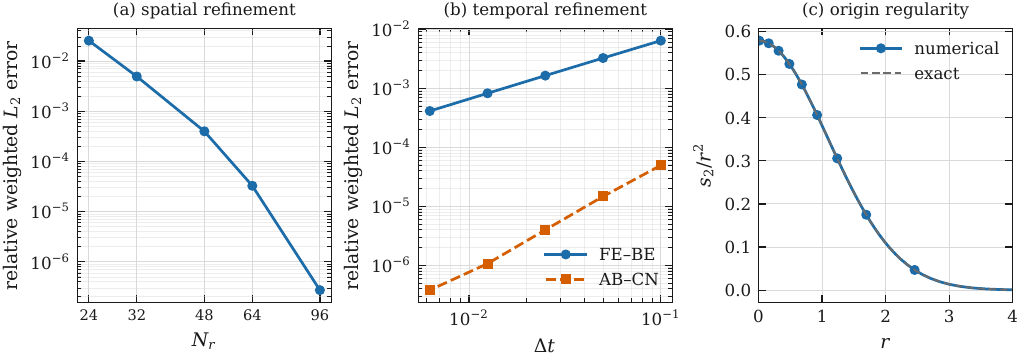}
\caption{Manufactured $m=2$ heat mode of Eq.~\eqref{eq:gaussian-m} at $\nu=0.1$, $\sigma_0=1$, $\ell=1$, $t_f=1$. (a) Relative weighted $L_2$ error against radial truncation. (b) Error against time step for the two integrators. (c) The origin-regularity quotient $s_2/r^2$ against radius, with the exact profile dashed.}
\label{fig:manufactured}
\end{figure}

For the manufactured mode (Fig.~\ref{fig:manufactured}) the spatial error falls from $2.6\times10^{-2}$ at $N_r=24$ to $2.7\times10^{-7}$ at $N_r=96$, and the origin-regularity quotient $s_2/r^2$ converges with it, its maximum relative deviation from the exact profile falling from $6.6\times10^{-3}$ to $2.4\times10^{-8}$ while remaining smooth and finite to the innermost node. Refining the time step instead recovers the same first- and second-order behavior as the Oseen sweep until the AB--CN error reaches the spatial error (Fig.~\ref{fig:manufactured}b). The azimuthal purity holds to round-off: the energy outside the $m=2$ harmonic stays below the double-precision floor at every resolution, and across the same sweep the azimuthal mean never exceeds $8.5\times10^{-17}$. That purity rules out a class of packing and redistribution errors.

\subsection{Example 2: reaction--diffusion front propagation}
\label{sec:ex2}

\paragraph{Problem}
The second example considers a localized population, chemical species or flame kernel that both diffuses and multiplies. Such a structure does not simply spread: after a transient it organizes into a front of fixed shape that advances at a selected speed, invading the surrounding medium indefinitely. This is the Fisher--Kolmogorov--Petrovsky--Piskunov problem \cite{fisher1937wave,kpp1937etude}, one of the most studied models in mathematical biology \cite{murray2002mathbio}. It is an apt test for an unbounded solver because the front never stops. On a truncated domain the calculation ends, or becomes contaminated, when the front reaches the outer boundary; on the mapped domain there is no boundary to reach, and the front can be followed until it outruns the resolved region instead.

\paragraph{Governing equations and initial conditions}
The problem is governed by
\begin{equation}
\frac{\partial u}{\partial t}=D\nabla^2u+\gamma u(1-u),
\label{eq:kpp}
\end{equation}
where $D$ denotes the diffusivity and $\gamma$ the linear reaction rate, on the unbounded plane, with $D=1$ and $\gamma=1/4$, so that the classical planar front speed is $c^*=2\sqrt{D\gamma}=1$ and the front width $\sqrt{D/\gamma}=2$ is comfortably resolved.

Two initial conditions are used. For the quantitative work the seed is a single axisymmetric Gaussian, $u(r,0)=A\exp(-r^2/2\sigma^2)$ with $\sigma=1$; for the qualitative figure it is three such Gaussians at radii $6$, $9$ and $11$, azimuths $132^\circ$, $0^\circ$ and $241^\circ$, and amplitudes $0.5$, $1$ and $1.5$. Equal seeds on an equilateral triangle would carry exact three-fold symmetry and thus excite only azimuthal orders $m=0,3,6,\dots$; giving each seed a different radius, azimuth and amplitude removes that degeneracy and puts energy in every order. No boundary condition is imposed at a finite radius. The far field ahead of the front is empty rather than held at a wall.

\begin{figure}[t]
\centering
\includegraphics[width=\textwidth]{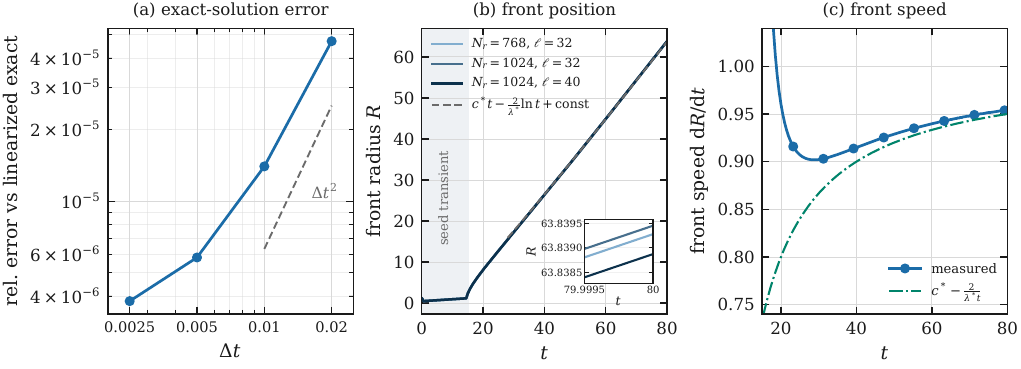}
\caption{Fisher--KPP validations, all for the single axisymmetric Gaussian seed. (a) Relative error against the exact linearized solution of Eq.~\eqref{eq:kpp-linear} as the time step is refined, with a second-order guide. (b) Front radius against time for two radial truncations and two map scales, which are indistinguishable at this scale, with the logarithmically corrected law; the shaded interval is the seed transient and the additive constant is fitted. (c) Instantaneous front speed against the logarithmically corrected estimate.}
\label{fig:kppval}
\end{figure}

\paragraph{Results and validation}
The quantitative work uses the single axisymmetric seed, for which two independent checks are available, one exact and one asymptotic. While $u\ll1$ the reaction is linear and Eq.~\eqref{eq:kpp} has the closed-form solution
\begin{equation}
u(r,t)=A\,\frac{\sigma^2}{\sigma^2+2Dt}\,
\mathrm{e}^{\gamma t}\exp\left[-\frac{r^2}{2(\sigma^2+2Dt)}\right],
\label{eq:kpp-linear}
\end{equation}
which tests the mapped diffusion operator, the reaction term, the semi-implicit split and the far-field treatment together against an exact answer. Taking $A=10^{-6}$ keeps the solution linear throughout, and at $t=20$ the relative weighted $L_2$ error is $1.401\times10^{-5}$ at $\Delta t=10^{-2}$. That figure is unchanged to five digits at $N_r=96$, $144$, $192$ and $256$, and thus the spatial error lies below it. Refining the time step gives $4.71\times10^{-5}$, $1.40\times10^{-5}$, $5.80\times10^{-6}$ and $3.81\times10^{-6}$, close to second order until it meets a floor near $3\times10^{-6}$ set by the quadratic reaction term that the linearized solution omits at this amplitude (Fig.~\ref{fig:kppval}a).

The nonlinear stage, computed from the same seed with $A=1$, is compared against front-speed theory. Figure~\ref{fig:kppval}b shows the measured front radius, defined as the largest radius at which the azimuthally averaged field exceeds one half. The front does not advance before $t\approx15$: as the seed spreads diffusively its peak falls below one half, and no half-level radius exists until the reaction lifts the core back above one half shortly before $t=15$, after which the radius rises sharply. That interval is shaded in the figure and excluded from the comparison. Thereafter the trajectory is insensitive to discretization to a remarkable degree: at $t=80$ the pair $N_r=768$ and $1024$ at $\ell=32$ and a third run at $\ell=40$ give $R=63.8393$, $63.8395$ and $63.8389$, agreeing to five significant figures across a change of truncation and of map scale, and thus any departure from theory is asymptotics rather than numerics.

The comparison is worth making at long times because the correction being tested shrinks as the run proceeds. In $N$ dimensions the front radius carries a logarithmic correction $R(t)=c^*t-[(N+2)/(2\lambda^*)]\ln t+O(1)$ with $\lambda^*=\sqrt{\gamma/D}$, of which the curvature contributes $(N-1)/(2\lambda^*)$ and the pulled-front mechanism the remaining $3/(2\lambda^*)$ \cite{roquejoffre2019sharp,buenzli2022curvature}; here it predicts $\mathrm{d}R/\mathrm{d}t=c^*-4/t$. At $t=32$ that correction is still $12.5\%$ of $c^*$ and the measured speed, $0.9037$, sits $0.0287$ above the predicted $0.8750$. Advancing to $t=80$ reduces the correction to $5.0\%$ of $c^*$ and the discrepancy to $0.0042$, a factor of about $7$, with the measured $0.9542$ against a predicted $0.9500$ (Fig.~\ref{fig:kppval}c). The selected speed therefore approaches $2\sqrt{D\gamma}$ from below along the predicted path.

\begin{figure}[t]
\centering
\includegraphics[width=\textwidth]{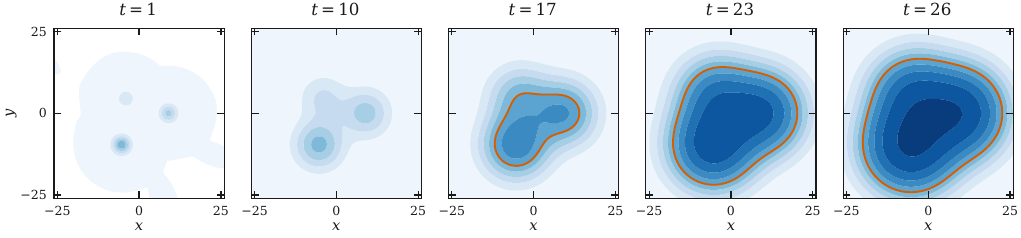}
\caption{Fisher--KPP front from three seeds of unequal strength and spacing on the unbounded plane, $D=1$ and $\gamma=1/4$. Shading is $u$ from $0$ to $1$ and the line marks $u=1/2$. The colonies grow and merge into one lobed, asymmetric front that continues to expand with no outer boundary to reach.}
\label{fig:kppsnap}
\end{figure}

With the front speed established on the axisymmetric seed, the three-seed initial condition of Fig.~\ref{fig:kppsnap} shows what the same solver does when the problem is not axisymmetric at all. Each seed first decays as diffusion outruns the reaction, then recovers and begins to propagate, the strongest establishing itself first. The fronts expand and merge into a single front, which then slowly loses the memory of its origins: curvature smooths away the indentations left where the fronts joined, while the offset of the composite from the axis persists. By $t=26$ the invaded region is one lobed patch, still visibly asymmetric and still growing, with no boundary for it to reach. Because the seeds share no symmetry, every azimuthal order is excited rather than one order in three.

\subsection{Example 3: vortex pair simulations}
\label{sec:ex3}

\paragraph{Problem}
The third example exercises the three-dimensional vector formulation with a pair of parallel columnar vortices. We consider two canonical cases. Vortices of the same sign rotate about their centroid and remain compact, making this configuration useful for a resolution study of a three-dimensional nonlinear calculation. Vortices of opposite sign instead propagate and are linearly unstable to the long-wavelength symmetric disturbance identified by Crow \cite{crow1970stability}. Its growth rate, most-amplified wavelength, and perturbation-plane inclination are predicted analytically and have been reproduced experimentally and numerically \cite{bristol2004cooperative,williamson2014vortexpairs,garten2001crow}.

\paragraph{Governing equations and initial conditions}
Both cases solve the incompressible Navier--Stokes equations in toroidal--poloidal form, advancing
\begin{equation}
\frac{\partial}{\partial t}(\psi,\chi)
=\mathcal{P}\!\left[-\boldsymbol{\omega}\times(\boldsymbol{U}+\boldsymbol{v})\right]
+\mathcal{L}(\psi,\chi),
\label{eq:tp-evolution}
\end{equation}
where $\mathcal{P}$ denotes toroidal--poloidal projection, $\boldsymbol{v}$ is the velocity field ($\mathcal{P}(\boldsymbol{v})=(\psi,\chi)$), $\boldsymbol{\omega}=\nabla\times\boldsymbol{v}$ is the vorticity, $\boldsymbol{U}$ is a uniform background flow, and $\mathcal{L}$ is the dissipation operator of Eq.~\eqref{eq:dissipation}. In both cases, the initial state is a superposition of two columnar vortices with circulations $\Gamma_1$ and $\Gamma_2$ at $(x,y)=(\pm b/2,0)$. Both cases use $\Delta t=10^{-2}$ unless stated otherwise.

For the co-rotating case, $\Gamma_1=\Gamma_2$ and each vortex has the $q$-vortex profile $V_\phi=(1-\mathrm{e}^{-r^2})/r$ and $V_z=q^{-1}\mathrm{e}^{-r^2}$. We take $b=4$, $q=1$, a uniform axial background $U_z=-0.5$, $\ell=4$, $L_z=2\pi$, and physical viscosity $\nu=10^{-4}$, giving a circulation Reynolds number $\Gamma/\nu=6.3\times10^4$.

For the counter-rotating case, $\Gamma_1=-\Gamma_2=2\pi$ and the vortices have Lamb--Oseen cores, $\omega\propto\mathrm{e}^{-r^2/a^2}$, of radius $a=0.5$ (i.e., $a/b=0.125$). There is no axial jet, and the axial period is the classical most-amplified wavelength, $L_z=8.6b$. This case uses the same $\nu=10^{-4}$ and hence the same $\Gamma/\nu=6.3\times10^4$, whereas the hyperviscosity $\nu_h$ is the only dissipation parameter varied between runs. All counter-rotating runs also enable the bounded filter of Section~\ref{sec:time}, with $\zeta_c=0.75$, $\tau=10^{-12}$, an amplitude saturating at $\epsilon=1$ and a relaxation factor of $0.25$. Without it, the high-mode tail grows from the start, leading to numerical blow-up. We seed the disturbance by displacing the two cores sinusoidally along $z$, mirror-symmetrically about $x=0$, in a plane inclined at $48^\circ$. The vortical field decays in the far field without an imposed boundary condition, and because the pair has zero total circulation, that decay is dipolar.

The background $\boldsymbol{U}$ supplies the uniform axial offset $U_z$ in the co-rotating case. In the counter-rotating case, it instead supplies a translation because the pair is not stationary: each vortex is advected by the other, and the pair translates at $\Gamma/(2\pi b)$. On a grid that concentrates resolution near $r=0$, the pair would drift out of the well-resolved region during the long integration required by the instability. We therefore integrate in the moving frame, adding a uniform background to the advecting velocity in physical space.

\begin{figure}[t]
\centering
\includegraphics[width=\textwidth]{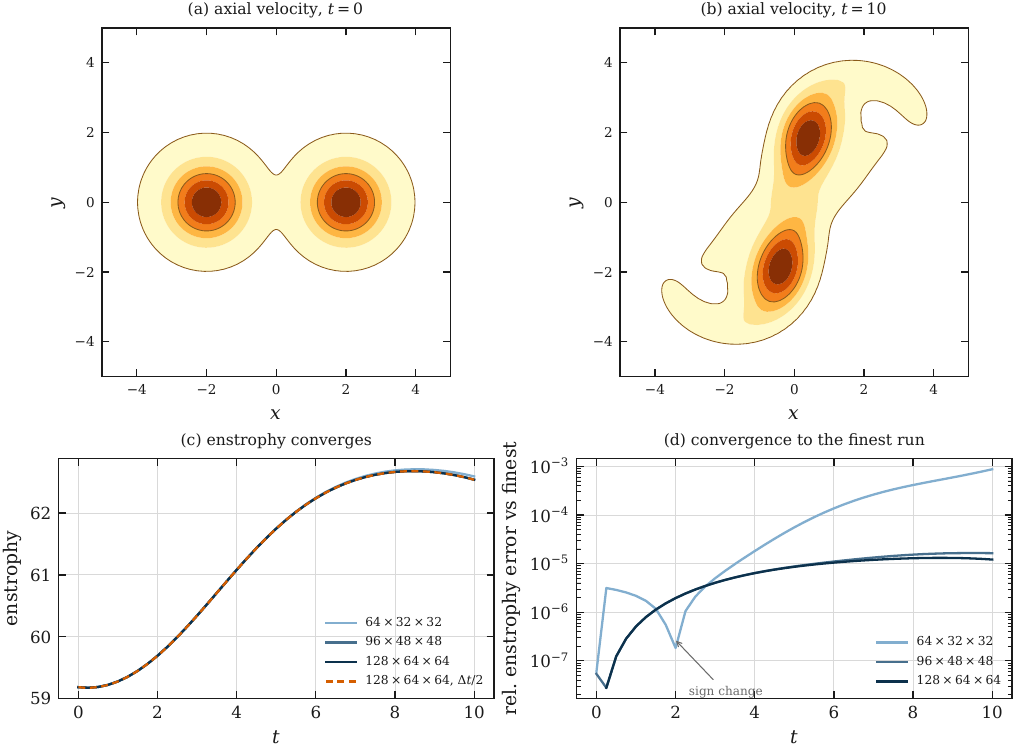}
\caption{Co-rotating $q$-vortex pair at $b=4$, $q=1$, $\ell=4$ and $\Gamma/\nu=6.3\times10^4$. (a, b) Axial velocity on the $z=0$ plane at $t=0$ and $t=10$, showing the cores orbiting their centroid. (c) Enstrophy history at three resolutions and at a halved time step, the last being the reference, indistinguishable on this scale. (d) Relative error in enstrophy against that reference.}
\label{fig:corot}
\end{figure}

\paragraph{Results and validation}
The co-rotating pair remains compact, and its enstrophy histories approach one another with refinement (Fig.~\ref{fig:corot}). The cores orbit their centroid, remain resolved, and shed only a weak filament by $t=10$. We use the finest ladder run ($128\times64\times64$ at half the time step) as the reference. At $t=10$, the relative enstrophy difference falls from $8.7\times10^{-4}$ at $(64,32,32)$ to $1.6\times10^{-5}$ at $(96,48,48)$ and $1.2\times10^{-5}$ at $(128,64,64)$. The last run uses the reference spatial grid at the larger time step, and hence its difference from the reference measures time-step sensitivity on that grid. The $(96,48,48)$ result is within a factor of two of this difference. A separate $200\times128\times128$ run returns $62.5397$ at $t=10$, agreeing with the $128\times64\times64$ entry at the same time step to six significant figures. The cusp in the coarsest curve near $t\approx1.9$ marks a change of sign: that run underestimates the enstrophy before the cusp and overestimates it afterward.

\begin{figure}[t]
\centering
\includegraphics[width=\textwidth]{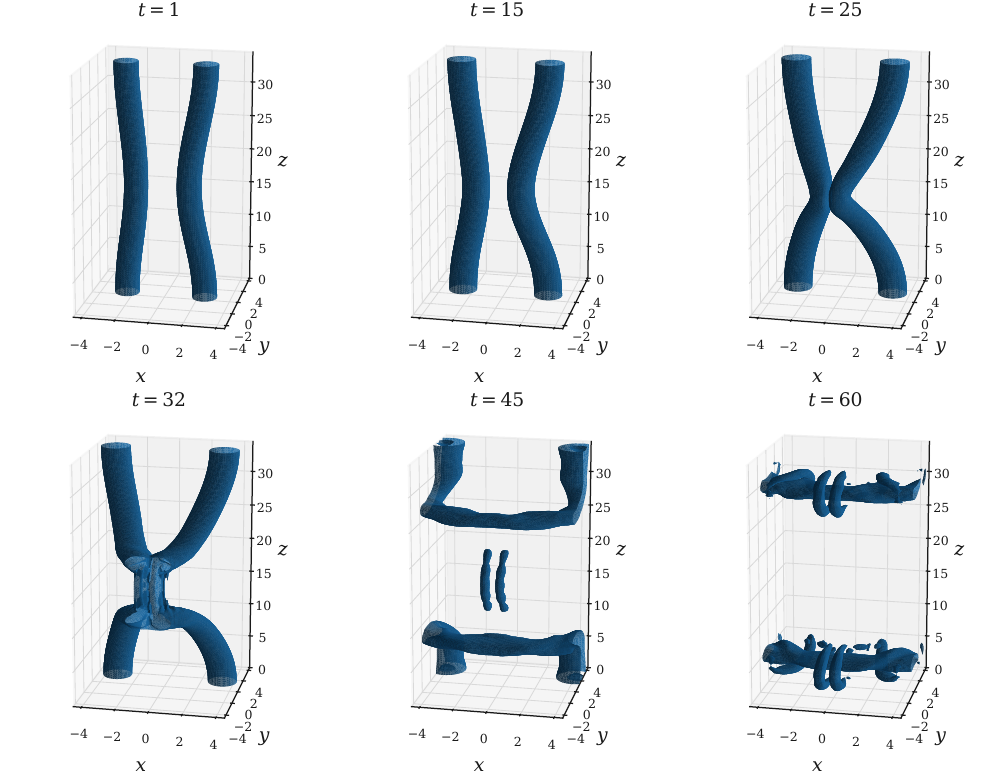}
\caption{Iso-surfaces of vorticity magnitude, $|\boldsymbol{\omega}|=1$, through the Crow instability of the counter-rotating pair with $b=4$, $a=0.5$, $L_z=8.6b$ and a $0.15b$ symmetric seed, computed on a $120\times96\times64$ grid. The axial coordinate spans exactly one Crow wavelength; the radial domain is unbounded and the window shown is $|x|,|y|\leq4.5$.}
\label{fig:crowsnaps}
\end{figure}

\begin{figure}[t]
\centering
\includegraphics[width=\textwidth]{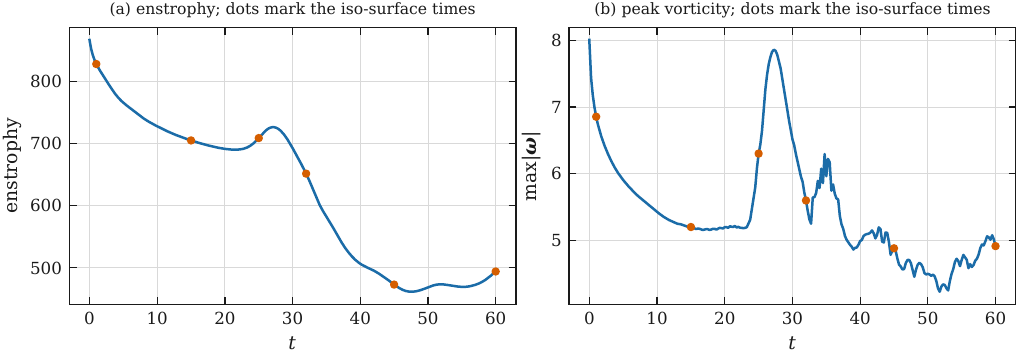}
\caption{Histories for the counter-rotating pair from the $176\times120\times96$ long run. (a) Enstrophy to $t=60$, with dots at the six times shown as iso-surfaces in Fig.~\ref{fig:crowsnaps}. (b) Peak vorticity magnitude.}
\label{fig:crowhist}
\end{figure}

The counter-rotating pair behaves differently. As Fig.~\ref{fig:crowsnaps} shows, the seeded wave grows; the tubes approach and meet at the mid-plane, reconnect into a bridge, close into a train of rings, and finally break down into smaller-scale structure. Figure~\ref{fig:crowhist} follows the integral diagnostics through this sequence. Enstrophy falls from $867$ while the pair is smooth, reaches a minimum of $461$ at $t=47.5$ as the tubes thin and reconnect, and rises to $494$ by $t=60$ as the rings form. The peak vorticity follows the same trend, bottoming at $4.23$ before recovering.

This demonstration extends to $t=60$, through reconnection and the formation of the ring train. The run continues beyond that point and does not fail, but following the complete small-scale breakdown is out of scope: the late fluctuations are violent, and only one calculation is available at this configuration, and thus we do not present them as converged. The long run uses $\nu_h=5\times10^{-7}$ on a $176\times120\times96$ grid rather than the smaller $\nu_h$ used for the growth-rate calculation, to maintain stability through the small-scale phase; the growth rate is not measured from it.

The linear stage of that sequence can be checked quantitatively against theory. A linearly unstable disturbance grows exponentially, its amplitude as $\mathrm{e}^{\sigma t}$ and hence its energy as $\mathrm{e}^{2\sigma t}$, and thus we extract the rate from the axial modal energy $E_k$ of the seeded fundamental and report it non-dimensionally, in units of $\Gamma/(2\pi b^2)$, as $\alpha=(\pi b^2/\Gamma)\,\mathrm{d}\ln E_1/\mathrm{d}t$, which for the present $\Gamma=2\pi$ reduces to $\alpha=\frac12 b^2\,\mathrm{d}\ln E_1/\mathrm{d}t$. We evaluate Crow's dispersion relation rather than his tabulated peak, since our core is Gaussian, whereas his calibration assumes a uniform one. Writing $\xi(\beta)=\beta K_1(\beta)$, $\eta(\beta)=\beta^2K_0(\beta)+\beta K_1(\beta)$ and $\varpi(\delta)=\frac12[(\cos\delta-1)/\delta^2+\sin\delta/\delta-\mathrm{Ci}(\delta)]$, in which $K_0$ and $K_1$ denote modified Bessel functions of the second kind and $\mathrm{Ci}$ the cosine integral, the symmetric mode satisfies \cite{crow1970stability}
\begin{equation}
\alpha_S^2=\left(1-\eta+\beta^2\varpi\right)\left(1+\xi-\beta^2\varpi\right),
\label{eq:crow-dispersion}
\end{equation}
where $\alpha_S$ is that mode's growth rate in the same units, $\beta=\kappa b$ and $\delta=\kappa d$, with $\kappa$ the axial wavenumber and $d$ the cutoff distance, which for the Lamb--Oseen core used here is $d=a\,\mathrm{e}^{(1-\gamma_E-\ln2)/2}\simeq0.8736\,a$, where $\gamma_E$ is the Euler--Mascheroni constant. Our implementation reproduces Crow's own results to better than $1\%$, returning $\beta_{\max}=0.7356$, $\alpha_{\max}=0.8275$, an inclination of $47.66^\circ$ and $\lambda_{\max}/b=8.542$ against his $0.73$, $0.83$, $48^\circ$ and $8.6$; evaluated at our $\beta=0.7306$ and $a/b=0.125$ it predicts $\alpha_{\mathrm{theory}}=0.7983$. The measurement uses a $128\times128\times32$ grid, a core displacement of $0.002b$ and a hyperviscosity $\nu_h=5\times10^{-8}$; the rate plateaus at $\alpha=0.8201$ over $t=33$--$60$, exceeding the target by $2.7\%$. Two sensitivity checks bound the measurement, both over the earlier window $t=10$--$18$ where every run overlaps: raising the seed amplitude tenfold, to $0.02b$, shifts the fitted rate by $0.3\%$, and refining the grid to $160\times128\times48$ shifts it by $0.32\%$. The amplitude pair was run on a coarser $96\times48\times32$ grid and at the larger $\nu_h$, and the finer grid was not carried into the plateau window, and thus neither check is a direct replication of the plateau itself; both nonetheless place the sensitivity well below the $2.7\%$ gap to theory. %

\begin{figure}[t]
\centering
\includegraphics[width=\textwidth]{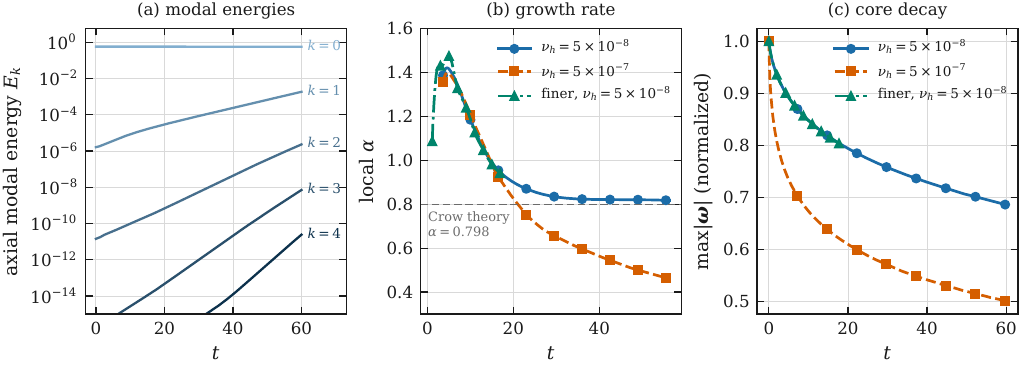}
\caption{Crow-instability growth-rate verification. (a) Axial modal energies; the fundamental grows exponentially while $k=2$--$4$, placed outside the unstable band by Eq.~\eqref{eq:crow-dispersion}, follow as nonlinear harmonics. (b) Local growth rate at two hyperviscosities and, at the smaller one, a finer grid, with the analytical target dashed. (c) Peak vorticity, showing the core diffusion responsible for the difference.}
\label{fig:crow}
\end{figure}

We turn last to an artifact of the numerical dissipation. Figure~\ref{fig:crow}a shows the axial modal energies from which the rate is taken: the fundamental grows exponentially, while the modes $k=2$--$4$, which Eq.~\eqref{eq:crow-dispersion} places outside the unstable band, grow only as its nonlinear harmonics rather than as independent instabilities. Figure~\ref{fig:crow}b shows the rate itself at two hyperviscosities. At $\nu_h=5\times10^{-7}$ it does not level off, falling monotonically through the predicted value and still decreasing at $\alpha\approx0.46$ by $t=60$; reducing $\nu_h$ tenfold produces the plateau quoted above, and a finer grid at that smaller coefficient tracks it over the interval the two share. Figure~\ref{fig:crow}c gives the reason. Peak vorticity falls by $49.9\%$ over $t=0$--$60$ at the larger coefficient against $31.4\%$ at the smaller, and since $\nu$ and the filter settings are identical in the two runs that difference is the powered dissipation acting on the vortex core itself, which is far larger than the smallest retained scales that the hyperviscosity is meant to damp. The tenfold larger coefficient therefore erodes the base flow on which the instability grows, depressing the measured rate steadily below the theoretical value, whereas the smaller one leaves the growth intact. 

These runs stay linear throughout, the fundamental reaching only $6\%$ of the base amplitude by $t=60$, and thus the cores are never stretched and the decay is purely diffusive; the non-monotonic peak vorticity of Fig.~\ref{fig:crowhist}b belongs to the large-amplitude run, where reconnection amplifies vorticity faster than dissipation removes it. The requirement is therefore to make the dissipation small enough that the measured rate stops moving, not to eliminate its effect on the core. 

Switching the powered dissipation off entirely, at unchanged $\nu$, fails after 1508 steps at $96\times48\times32$ and after only 427 at $128\times128\times32$, the finer grid failing sooner even though its initial condition is represented to $7\times10^{-13}$. The cause is not under-resolution of the cores but the mapped grid itself: the arc length $r\Delta\phi$ collapses as $r\rightarrow0$, and thus the effective azimuthal wavenumber near the axis is very large and explicit advection there imposes a time-step restriction tighter than a core-scale estimate suggests. That time step is viable only because the powered dissipation damps those near-axis modes, and users reducing $\nu_h$ must reduce $\Delta t$ with it.

\section{Conclusions}
\label{sec:conclusions}

We have presented \mlegs, an open-source code package for time-dependent partial differential equations with one radially unbounded coordinate. The mapped associated Legendre basis ties the near-origin behavior of each harmonic to its azimuthal order and carries the far-field decay analytically, and thus no boundary condition is imposed at a finite radius or at the axis. Relative to the original prototype of this numerical lineage, the contribution is an engineered package: a module and submodule hierarchy that can be revised one piece at a time, and a distributed spectral scalar that presents ready-made fields and operators to a driver in place of array indices, with transposes carried out through MPI subarray datatypes and the basis tables built in parallel. On the supercomputing system used here, the package scales across four compute nodes.

The three examples establish accuracy and, in several places, the limits of what the formulation can be asked to deliver. Spatial convergence separates cleanly into the two regimes the basis predicts, and the dependence on the map scale is strong enough that we regard $\ell$ as a resolution parameter to be swept rather than a value inherited from an example input. The two semi-implicit integrators recover their design orders, and the same study shows that the circulation and the second radial moment stop responding to the time step once the spatial error limits them, an ordering a purely temporal study would not reveal; a manufactured non-axisymmetric mode shows azimuthal leakage at round-off level at every resolution. The reaction--diffusion front reproduces its exact linearized solution and follows a trajectory converged to five significant figures, propagating at the classical speed to within the known logarithmic correction; it also demonstrates the practical value of an unbounded representation for a structure that never stops moving outward. The vortex-pair example reproduces the Crow growth rate to $2.7\%$ and carries the pair through reconnection and ring formation. It also shows what the numerical dissipation costs if it is left too large: at a tenfold larger $\nu_h$ the powered term reaches the vortex cores themselves, which are far larger than the smallest retained scales it is meant to damp, and the measured growth rate does not level off.

We anticipate three extensions. First, a wavenumber-aware hyperviscosity would replace the single global $\nu_h$ applied with one fixed power. In the counter-rotating runs of Example 3, the bounded filter maintains stability only when it acts together with the powered dissipation. A coefficient that responds to the growth of the high-mode tail and varies across wavenumbers would damp only the modes that need it. Such a scheme would also help in the radial direction, where the periodic two-thirds mask does not apply. A radial margin there requires setting $N_r^{c}$ below $N_r$, which costs resolution or grid points. In addition, energy already folded from unresolved degrees into retained ones cannot be removed by filtering afterwards. Second, dynamic time stepping would address the time-step limit, which here is set by the near-axis azimuthal spacing rather than by the cores. This limit is also coupled to the dissipation. As a result, a step chosen for the initial condition is either wasteful later or unsafe, and a user who reduces $\nu_h$ must currently reduce $\Delta t$ along with it. Adapting the step to the measured near-axis advective limit would break that coupling and remove a pitfall that no diagnostic presently warns about. Finally, an adaptive map scale would help because $\ell$ is fixed for a run and chosen by hand, although it is the strongest single control on accuracy. At $N_r=64$, the Gaussian representation error spans six orders of magnitude between $\ell=0.5$ and $\ell=2$. A structure that expands or migrates outward, such as the reaction--diffusion front, also eventually moves beyond the region that $\ell$ resolves. Rescaling or translating the map during a run, as has been developed for other unbounded bases \cite{xia2021scaling}, would let the resolved region follow the solution and remove the manual parameter search of the present study.

\section*{CRediT authorship contribution statement}
\textbf{Sangjoon Lee:} Conceptualization, Methodology, Software, Validation, Formal analysis, Investigation, Data curation, Visualization, Writing -- original draft, Writing -- review \& editing. \textbf{Jinge Wang:} Conceptualization, Methodology, Software, Validation, Investigation, Writing -- review \& editing.

\section*{Data availability}
\mlegs\ is publicly open at \url{https://github.com/UCBCFD/MLegS}, comprising the source code, documentation, and sample programs. The versioned software releases, including v1.1.4 reported here, are separately indexed \cite{lee2026mlegs}. The data and scripts behind the figures and tables are available from the corresponding author on request.

\section*{Acknowledgements}
The authors thank the UC Berkeley CFD laboratory, directed by Professor Emeritus Philip S.\ Marcus, whose archetypal F77 implementation of the mapped Legendre spectral method serves as the foundation on which \mlegs\ has been built. For scaling tests, this work used Anvil at Purdue University through allocation PHY250071 from the Advanced Cyberinfrastructure Coordination Ecosystem: Services \& Support (ACCESS) program, which is supported by U.S.\ National Science Foundation grants \#2138259, \#2138286, \#2138307, \#2137603, and \#2138296.

\bibliographystyle{elsarticle-num}
\bibliography{ref}

\end{document}